# Heterogeneous-free narrow linewidth semiconductor laser with optical injection locking

## Author Information

Xiao Sun[1*], Zhibo Li[1], Yiming Sun[1], Yupei Wang[1], Jue Wang[1], John H. Marsh[1], Stephen. J. Sweeney[1], Anthony E. Kelly[1], Lianping Hou[1*]

1. James Watt School of Engineering, University of Glasgow, Glasgow, G12 8QQ, U.K.

* Correspondence: Xiao Sun (Xiao.Sun@glasgow.ac.uk)
James Watt School of Engineering, University of Glasgow
Glasgow, G12 8QQ, U.K
Email: Xiao.Sun@glasgow.ac.uk

* Correspondence: Lianping Hou (Lianping Hou @glasgow.ac.uk)
James Watt School of Engineering, University of Glasgow
Glasgow, G12 8QQ, U.K
Email: Lianping.Hou@glasgow.ac.uk

## Abstract

Narrow-linewidth lasers are indispensable for coherent optical systems, including communications, metrology, and sensing. Although compact semiconductor lasers with narrow linewidths and low noise have been demonstrated, their spectral purity typically relies on hybrid or heterogeneous external-cavity feedback. Here, we present a theoretical and experimental demonstration of a heterogeneous-free optical injection locking (HF-OIL) semiconductor laser. By integrating a topological interface state extended (TISE) laser with a micro-ring resonator (MRR) on an AlGaInAs multiple-quantum-well platform, we achieve monolithic photon injection and phase locking, thereby reducing the optical linewidth. We fabricated and characterized a 1550 nm sidewall HF-OIL laser, achieving stable single-mode operation over a broad current range (65–300 mA) and a side-mode suppression ratio (SMSR) > 50 dB. Under injection locking, the device's Voigt-fitted linewidth narrowed from >1.7 MHz (free-running) to 4.2 kHz, representing a three-order-of-magnitude improvement over conventional distributed-feedback lasers. The intrinsic linewidth of 1.4 kHz is measured by correlated delayed self-heterodyne frequency noise power spectrum density (FN-PSD) method. Moreover, the HF-OIL laser demonstrated high phase stability and the ability to transition from a random-phased to a phase-locked state. These results underscore the potential of HF-OIL lasers in advancing coherent optical communications and phase encoders in quantum key distribution (QKD) systems.

## Introduction

Lasers with high-frequency spectral purity are critical for numerous applications, including sensing and spectroscopy [1,2], optical clocks[3,4], LIDAR[5,6], and microwave photonic devices[7-9]. Moreover, the demand for low frequency and amplitude noise is paramount in high-data-rate optical communications, particularly in coherent systems[10-14]. Although conventional all-solid-state[15] and fiber lasers [16] offer excellent performance, their reliance on discrete, bulky components and low energy conversion efficiency hampers on-chip integration. Semiconductor lasers, by contrast, are more amenable to integration but inherently exhibit linewidths in the MHz range[17], primarily due to their short cavity lengths and limited cavity quality factors is not suitable for the high-speed modulation 16-QAM or 64-QAM requirement (linewidth < 100 kHz).

To address these limitations, prior research has demonstrated that coupling a distributed feedback (DFB) semiconductor laser with a high-Q external optical cavity can create a self-injection-locked (SIL) system. The resulting narrow-band filtering feedback from the external cavity significantly suppresses the laser's phase and frequency noise, thus enhancing its frequency stability. Previous studies have explored various external cavities, such as Fabry-Perot (F-P) etalons[18,19], fiber ring resonators[20], and whispering-gallery-mode resonators (WGMRs)[17], though these often rely on large-scale fiber-based systems.

For on-chip integration, a common strategy involves hybrid integration of a DFB laser with a micro-ring resonator (MRR) on silicon on insulator (SOI)[21-23], lithium niobate on insulator(LNOI)[24-26], and $Si_3N_4$-based platforms. $Si_3N_4$[27-31] is widely used to make the external optical cavity due to its low loss, but its footprint is relatively large due to the small refractive index contrast of the $Si_3N_4$ waveguide. Moreover, phase tuning of $Si_3N_4$ is mainly based on the thermo-optic effect, which limits the tuning speed to tens of kilohertz[32]. Rayleigh backscattering in the high-Q (~$10^7$) MRR can reduce the laser linewidth to the sub-kHz level. However, under weak feedback conditions (typically <−20 dB), SIL is highly sensitive to the feedback phase and external perturbations. Moreover, the intensity of Rayleigh backscattered light re-injected into the DFB laser is limited by the coupling efficiency between the DFB device and the external cavity. This coupling is fundamentally limited by the challenges of heterogeneously integrating III-V materials onto silicon platforms. Specifically, achieving effective SIL and external cavity laser (ECL) designs requires high-efficiency coupling components, such as spot size converters (SSCs). Moreover, heterogeneous integration complicates the testing of individual components before their integration into complex systems. This necessitates stringent process control to ensure high yield, while also increasing fabrication complexity and cost.

Sideband Optical injection locking (SOIL) is a technique used to synchronize the frequency and phase of a "slave" laser with an external "master" laser, leveraging photon-photon interactions within the slave cavity[33]. Recently, monolithic SOIL (MSOIL) lasers—consisting of two back-to-back distributed feedback (DFB) lasers integrated on chips—have been proposed for high-speed optical communication[34-36], wavelength proportionality[37], and long-distance ranging[38]. Their adjustable phase characteristics further suggest potential applications in QKD systems[39]. However, in the back-to-back MSOIL configuration, no compact on-chip optical isolation exists between the master and slave lasers. As a result, the slave's output light can be reinjected into the master, leading to four-wave mixing and destabilizing the injection-locked state, thus narrowing the operating current range over which stable locking can be maintained[40]. In addition, the cavity quality factor of each sub-laser remains unchanged in this arrangement, thereby limiting the extent of achievable linewidth narrowing in injection locking. Previous research indicates that back-to-back MSOIL lasers maintain a linewidth of 14.88 MHz[38] and 2.7 MHz[35], comparable to that of conventional DFB lasers. Although injecting an external radio frequency (RF) signal can reduce the laser linewidth to the sub-kHz range[38,41], this method requires a low-noise, high-frequency RF source, which limits its practical feasibility.

In this paper, we explore the integration of SIL concepts into the design of monolithic SOIL laser systems. We propose and experimentally demonstrate a heterogeneous-free optical injection locking (HF-OIL) semiconductor laser configuration by monolithically integrating a topological interface state extended (TISE) laser and an MRR on an AlGaInAs multiple-quantum-well (MQW) platform. The TISE laser configuration establishes a topological interface state at the cavity center, leading to a more uniform photon density distribution and enhancing the optical coupling between the laser and the MRR.

We fabricated a sidewall HF-OIL laser with a single-mode locking operation DFB current range from threshold 65 mA to 300 mA, and a side mode suppression ratio (SMSR) above 50 dB even without anti-reflection (AR) facet coatings. Under injection locking, the intrinsic linewidth narrows significantly from 2 MHz to 4.2 kHz compared to the free-running state. The footprint of HF-OIL laser is only about 1000 × 0.2 μm$^2$, much smaller than the hybrid Si$_3$N$_4$ or SOI SIL/ECL laser. We also present a proof-of-concept demonstration using the coherent-one-way (COW) protocol to modulate the HF-OIL laser, illustrating its superior phase stability. Furthermore, a phase-switching modulation experiment confirms the device's ability to transition from a random-phase state to a phase-locked state, effectively toggling between locked and unlocked conditions. This HF-OIL system achieves superior single-mode, small footprint, narrow-linewidth performance without relying on a hybrid external cavity. Its fabrication is also simplified, requiring only a single metalorganic vapor-phase epitaxy (MOVPE) growth and one inductively coupled plasma (ICP) dry etching step. These results highlight the potential of the HF-OIL laser for integrated photonic platforms, offering improved stability, reduced complexity, and enhanced performance for coherent optical systems and phase encoder ability in QKD systems.

## Results

### Design of the HF-OIL laser

The traditional π-phase shifted uniform Bragg grating is shown in Fig. 1a. This structure consists of two Bragg reflection mirrors with a π-phase shift between them. Figure 1b presents the structure of the TISE laser, which includes left and right reflection uniform grating mirrors ($L_{LG}$ and $L_{RG}$) with inversion symmetry, and a central topological interface state extending (TISE) cavity ($L_M$). Here, $Λ_M$ is the lateral modulation period and $\Lambda$ denotes the cell period. The modulation process in the TISE cavity is shown in Fig. 1c, where $\Lambda$ is the Bragg grating period, $n_1$ and $n_2$ refer to two different refractive index segments. The positions of the $n_1$ and $n_2$ are changed with a shift step $\Delta\Lambda$ while maintaining a constant $\Lambda$ during each modulation cycle. Within the modulation period $\Lambda_M$, an alignment factor, $\Delta\Lambda$ quantifies the adjustment of the phase shift step across each sampling period $\Lambda_M$. After a lateral modulation period $\Lambda_M$, the cell returns to the initial state, resulting in $\Delta\Lambda = \Lambda$. For a lateral modulation period $\Lambda_M$ with $n^{th}$ modulation, it holds that $\Lambda_M = n\Lambda$. The TISE cavity contains $N$ number of $\Lambda_M$ and the total length of TISE ($L_M$) is equal to $N\Lambda_m$. Figure 1d illustrates, from left to right, the calculated 1D band structures of the left grating, the TISE cavity, and the right grating. Both the left and right gratings exhibit two Zak phase inversion centers, while the energy band in the TISE cavity is degenerate.

Figure 1e illustrates the structure of the HF-OIL laser, which integrates a TISE laser with an MRR on an AlGaInAs MQW platform which the gain peak center is around 1560 nm (see Supplementary Part A). The TISE laser, measuring 1000 μm in total length, comprises symmetrically inverted left and right reflective grating mirrors (450 μm each) flanking a central TISE cavity (100 μm) as shown in Figure 1f. The band-degenerated TISE cavity provides topological protection of the phase transition between the left and right gratings, thereby enabling the topological interface state (TIS) mode to extend uniformly across the cavity's center.

At the cavity midpoint, a 150 μm radius MRR is integrated close to the TISE cavity. Depending on our calculation result (see Supplementary Part A), for $R > 150$μm, the increase in the $Q$ factor is minimal, further increasing $R$ does not significantly improve the linewidth narrowing effect. As light propagates through the TISE laser, a portion of light is coupled into the MRR. After resonant circulation within the MRR, this light is re-injected into the TISE laser, achieving optical injection locking. Due to the structural asymmetry, the MRR can couple simultaneously to both left-propagating (green arrow) and right-propagating (red arrow) modes in the TISE laser cavity. Figures 1g and 1h show the transmission spectrum and normalized photon distribution of the TISE laser, compared against a conventional π-phase shift DFB laser. The results indicate that the TISE cavity effectively suppresses side modes and produces a more uniform field distribution at the waveguide center. This uniformity enhances the coupling efficiency between the TISE laser and the MRR. Figures 1i show the transmission spectrum from the MRR to the TISE laser and the quality factor $Q = 1.1 \times 10^5$. The efficiency of injection locking can be described by an equation derived for the reduction of the close-in linewidth of the laser (see Supplementary Part B).

$$\delta \approx \frac{P_r}{P}(1+\alpha_H^2)\left(\frac{Q_{MRR}}{Q_{LD}}\right)^2 \quad (1)$$

Where $P_r$ is the feedback power from the MRR, and $P$ is the TISE laser power. $Q_{LD}$ represents the quality factor of laser, while $Q_{MRR}$ denotes the quality factor of MRR, $\alpha_H$ is the linewidth enhancement factor. By substituting reasonable numbers in equation (1) for $P_r/P = 0.03$, $\alpha_H = 3$, $Q_{MRR} = 1.1 \times 10^5$, and $Q_{LD} = 4 \times 10^3$, we find that the laser linewidth can be improved by a factor of 225. The locking range is defined as the frequency range over which the TISE laser emission injection locks to the MRR resonance and follows the expression:

$$\Delta f_{lock} \approx \sqrt{\frac{P_r}{P}(1+\alpha_H^2)} \frac{f_0}{Q_{LD}} \quad (2)$$

Where $f_0$ represents centre frequency of the laser. The theoretically estimated locking range $\Delta f_{lock} \approx 26$ GHz.

Experiment

The devices were fabricated on an AlGaInAs/InP epitaxial structure featuring five quantum wells (QWs) and six quantum barriers, with a QW confinement factor of 5%[42]. The room-temperature photoluminescence (PL) wavelength of the QWs was set to 1530 nm. A scanning electron microscope (SEM) image of the HF-OIL laser is shown in Fig. 2a. The grating ridge waveguide was designed with a width of 2.0 μm, a sidewall corrugation depth of 300 nm, and a ridge height of 2.0 μm. The minimum gap between the MRR and the TISE laser was maintained at 300 nm.

The laser chip was mounted on a thermoelectric cooler (TEC) set to 20°C to mitigate long-term drift. The MRR is also carrier injected to reduce the internal absorption loss of the MQW. Figure 2b shows the typical current–power ($I_{TISE}$-$P$) characteristics of the TISE laser under different MRR injection currents ($I_{MRR}$). The threshold current is approximately 65 mA, and the slope efficiency is about 0.07 W/A. Notably, varying $I_{MRR}$ has only a minor influence on the TISE laser's output power. At $I_{MRR}$ = 300 mA, the saturation power is slightly reduced compared to $I_{MRR}$ = 150 mA, likely due to increased cavity temperature resulting from the higher injection current. The laser is still able to deliver power up to 11 mW. Figure 2c compares the optical spectra of the laser in its injection-locked and free-running states. Under injection locking, the spectral linewidth is notably narrowed, and the side-mode suppression ratio (SMSR) exceeds 50 dB. To examine the locking state range, we fixed $I_{TISE}$ at 200 mA and varied $I_{MRR}$ from 100 mA to 300 mA, as shown in Figure 2d. Once $I_{MRR}$ surpasses 171 mA, the spectrum transitions from a broad (free running) to a narrow (locked) linewidth, defining a locking range from 171 mA to 300 mA. The current-induced wavelength redshift coefficient (ACWRC) of the $I_{MRR}$ is found at 0.01022 nm/mA Figures 2e–f present 2D spectral maps of the laser for $I_{TISE}$ values ranging from its threshold to 300 mA, at $I_{MRR}$ = 0 mA, 150 mA, and 300 mA. The ACWRC for the $I_{TISE}$ is 0.01038 nm/mA, indicating that the wavelength redshift coefficient for TISE laser injection (with the $I_{MRR}$ fixed) closely matches that of MRR laser injection (with the $I_{TISE}$ fixed). As a result, the TISE laser and MRR exhibit excellent wavelength tuning synchronization, enabling a wide injection-locked range. These results confirm that the HF-OIL laser remains in the injection-locked state over a wide current range, maintaining stable locking for $I_{TISE}$ from threshold to 300 mA when $I_{MRR}$ = 300 mA.

Then we investigated the coherence properties of the HF-OIL laser. The laser linewidth was measured using a fiber-delayed, non-zero frequency delayed self-heterodyne (DSH) method, employing a 25-km single-mode fiber delay and an 80-MHz acoustic-optic modulator (AOM), as shown in Fig. 3a. A comparison of the free-running and injection-locked states is presented in Fig. 3b, revealing a pronounced linewidth narrowing under injection locking. Voigt fitting of the spectra is shown in Fig. 3c. Since the Gaussian noise has a more prominent effect on broadening the spectral line near the center [43], the intrinsic Lorentzian linewidth is estimated by measuring the –20 dB bandwidth and applying a $2\sqrt{99}$ division factor to reduce the effect of Gaussian noise[44]. Stable injection locking is obtained with an intrinsic Lorentzian linewidth of 4.2 kHz. We further examined the linewidth variation as a function of $I_{TISE}$ and compared it to the free-running state, as illustrated in Fig. 3d. Under free-running conditions, the intrinsic linewidth ranged from 1.7 MHz to 3.6 MHz, whereas injection locking reduced it to 4.2 kHz to 13 kHz—an improvement by a factor greater than 250, which is corresponded to the calculated reduction of the close-in linewidth factor. The linewidth is < 8 kHz within the $I_{TISE}$ range of 140 - 260 mA. Figure 3e shows the linewidth reduction as $I_{MRR}$ increases, with the laser transitioning from the free running to the locked state.

To confirm the linewidth, we employed the correlated delayed self-heterodyne frequency noise power spectrum density (FN-PSD) method (details in Methods). The recorded frequency noise is displayed in Fig. 3f. A white noise floor of approximately 446 Hz²/Hz is observed, which corresponds to an intrinsic linewidth (Lorentzian linewidth) of 1.4 kHz for the laser. This further confirms the narrow-linewidth performance of the laser in the locked state, in contrast to the free-running state, where the intrinsic linewidth is 2.6 MHz.

We evaluated the coherence of successive pulses in a pulse train under two optical injection conditions: a continuous wave (CW) MRR and a directly modulated MRR. The measurement setup, depicted in Fig. 4a, is based on an asymmetric Mach–Zehnder interferometer (AMZI). A thermal phase modulator (PM) in the short arm controls the relative phase between the two arms, while an acoustic-optic modulator (AOM) introduces an 80 MHz frequency shift. This shift facilitates the differentiation of low-frequency electronic noise from the beat signal on the PD, thus improving measurement accuracy and signal-to-noise ratio.

The TISE laser is driven by a 5 MHz RF signal with voltage of peak to peak at 400 mV, and the AMZI is configured with a 200 ns delay between its short and long arms—matching the 200 ns interval between consecutive pulses at 5 MHz (Fig. 4b). Under the CW-locked MRR condition, as illustrated in Fig. 4c, each pulse inherits the coherence from the locking process, and the phase difference between consecutive pulses remains fixed. This behavior closely mimics the saturation regime of the COW protocol. Conversely, when the MRR is directly modulated and the TISE laser operates in CW mode, the HF-OIL laser cycles between phase-locked and random-phase states, causing the phase of consecutive pulses to vary and no longer remain fixed.

## Discussion

The concept of injection locking—recently employed in narrow-linewidth semiconductor lasers through heterogeneous integration of III–V gain media and ultra-high-Q silicon-based microresonators—has been extensively explored for coherent optical systems. In contrast to heterogeneous approaches, monolithic integration unifies the material platform, reducing the device footprint, fabrication complexity and cost. However, most monolithic semiconductor lasers exhibit linewidth on the order of megahertz.

In this work, we present a heterogeneous-free HF-OIL laser fabricated on an InP platform. This device combines an MRR with a quality factor of $10^5$ and a novel TISE laser. The TISE laser employs a mode-extension cavity that expands the fundamental optical mode from the cavity center to its edges, producing a uniform photon density distribution at the cavity center and thereby enhancing its coupling efficiency to the MRR.

Our measurements show that the injection-locked HF-OIL laser achieves an Voigt-fitted linewidth of 4.2 kHz, FN-PSD linewidth 1.4 kHz more than 200 times narrower than its free-running state. Table 1 compares the performance of our HF-OIL laser with various hybrid and monolithic III-V narrow-linewidth semiconductor lasers. Notably, our device achieves a significantly narrower linewidth than typical monolithic III–V lasers, with performance on par with hybrid self-injection-locked and external-cavity lasers. The fabrication process of HF-OIL laser is also the simplest, with only one step of MOVPE epitaxial growth and one step ICP dry etch. Additionally, our experiments demonstrate the intrinsic phase-adjustment capability of the HF-OIL laser, highlighting its potential for both quantum and conventional coherent communication systems. This intrinsic phase control also enables the seamless combination of QKD and high-bandwidth data transmission, underscoring the versatility and promise of the HF-OIL laser platform.

**Table 1** Performance comparison of hybrid integrated SIL/ECL lasers and other narrow-linewidth DFB/DBR lasers in the 1550 nm.

| Reference | Structure | Power (mW) | SMSR (dB) | Linewidth (kHz) |
|---|---|---|---|---|
| 22 (2018) | Si-Hybrid ECL | 11 | > 46 | 37 (FN-PSD) |
| 45 (2021) | Si-Hybrid ECL | > 2 | > 40 | 105 (FN-PSD) |
| 46 (2024) | Si-Hybrid ECL | 16.4 | > 45 | 2.79 (DSH) |
| 21 (2022) | Si-Hybrid filter/SIL | > 2 | > 40 | 27 (FN-PSD) |
| 23 (2023) | Si-Hybrid ECL | 76 | > 50 | 12 (FN-PSD) |
| 47 (2020) | SiN-Hybrid ECL | 0.5 | > 55 | 4 (FN-PSD) |
| 48 (2021) | SiN-Hybrid SIL | 37.9 | 56 | 3 (FN-PSD) |
| 25 (2022) | LNOI-Hybrid ECL | 3.7 | > 50 | 15 (DSH) / 11.3 (FN-PSD) |
| 49 (2024) | LNOI-Hybrid SIL | 0.74 | / | 45.5 (DSH) |
| 50 (2024) | LNOI-Hybrid SIL | 3.18 | 60 | 2.5 (FN-PSD) |
| 51 (2015) | InP-DBR Butt joint | > 120 | > 50 | 70 (FN-PSD) |
| 52 (2022) | InP-QW DFB | 45 | 58.7 | 63 (DSH) |
| 53 (2022) | InP-DBR Butt joint | 18 | 54 | 45 (DSH) /10 (FN-PSD) |
| 54 (2019) | InP-QW DFB | / | / | 100 (DSH) |
| This Work | InP-QW HF-OIL | 11 | > 50 | 4.2 (DSH) /1.4 (FN-PSD) |

## Methods

### Device fabrication

The fabrication procedure for the HF-OIL laser can be found in Supplementary Part C. The wafer was grown on an InP substrate using MOVPE. The room temperature photoluminescence (PL) peak of the QWs was located at a wavelength of 1530 nm. The TISE laser sidewall grating and MRR waveguide was defined by electron-beam lithography (EBL) on an EBPG5200 E-beam system, with negative-tone Hydrogen Silsesquioxane (HSQ) acting as both the EBL resist and a hard mask for ICP dry etching using a $Cl_2$/$CH_4$/$H_2$ gas mixture in an Oxford PlasmaPro 300 system. Subsequent steps included PECVD deposition of $SiO_2$ (Oxford PlasmaPro 100 PECVD), application of HSQ passivation layers, $SiO_2$ window opening, P-contact deposition, substrate thinning, and N-contact deposition, all performed using conventional laser diode fabrication techniques. These two micro electrodes only require one step of lithography together with the large area electrode pads. Scanning electron microscopy images were acquired using a Hitachi SU8240 scanning electron microscope operating at 100 kV.

### Measurement setup

The laser was driven by a continuous-wave current source (Newport Model 8000). An isolator was placed at the laser output. The light passed through the isolator and entered coupler 1, which had a splitting ratio of 50:50, dividing the light into two paths. One path included a time-delay fiber and a polarization controller (FPC032) , while the other path included an 80-MHz acoustic-optic modulator (Aerodiode 1550-AOM). These two paths were then combined in coupler 2 and split again into two paths: one leading to a photodetector (Thorlabs DET08CFC) and the other to an optical spectrum analyzer (Agilent 86140B). Both the current driver and the electronic spectrum analyzer (Keysight N9000B) were controlled by a computer via the general-purpose interface bus (GPIB) interface using LabVIEW software. For the correlated delayed self-heterodyne FN-PSD method, the experimental setup is similar to the one shown in Fig. 3a, with the key difference being a 40 m long fiber line to delay the optical path.

The phase encoding experiment is conducted using the setup shown in Fig. 4a. A phase modulator (Thorlabs LNP6118) maintains the phase difference between the two AMZI paths, while the signal is analyzed with a real-time oscilloscope (Keysight UXR0334A).

## Data Availability

The data that support the findings of this study are available from the corresponding author upon reasonable request.

## Acknowledgment

The authors would like to thank the staff of the James Watt Nanofabrication Centre (JWNC) at the University of Glasgow for their help in fabricating the devices. The author sincerely thanks Rachel love in the assistance in JWNC and the Critical Technologies Accelerator for support of this research.

**Fig. 1: Design of HF-OIL semiconductor laser**

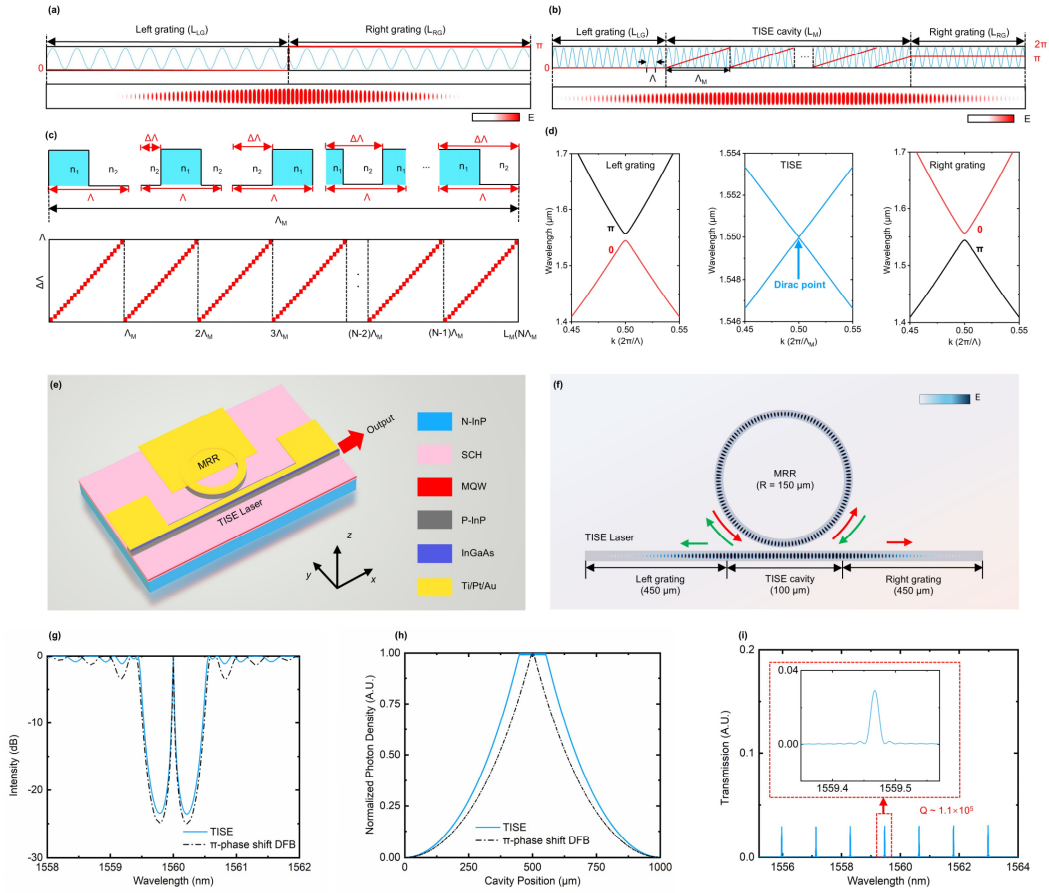

Structure and electrical field distribution of **a** traditional π-phase-shifted grating structure and **b** topological interface states extending (TISE) structure. **c** Modulation processing in TISE cavity including all the modulate periods. **d** Optical band structures of the left reflection grating ($L_{LG}$), topological interface states extending (TISE) grating ($L_M$), and right reflection grating ($L_{RG}$). The left-side and right-side gratings have distinct Zak phases and the TISE range to form a Dirac cone dispersion at $k = 0.5$. **e** Schematic of the HF-OIL laser. **f** Schematic with structure dimension and electrical field distribution. **g** transmission spectrum and **h** normalized photon distribution in the TISE laser compared to the conventional π-phase shift DFB laser. **i** Transmission spectrum from the TISE laser to MRR.

**Fig. 2: Device and spectrum narrowing**

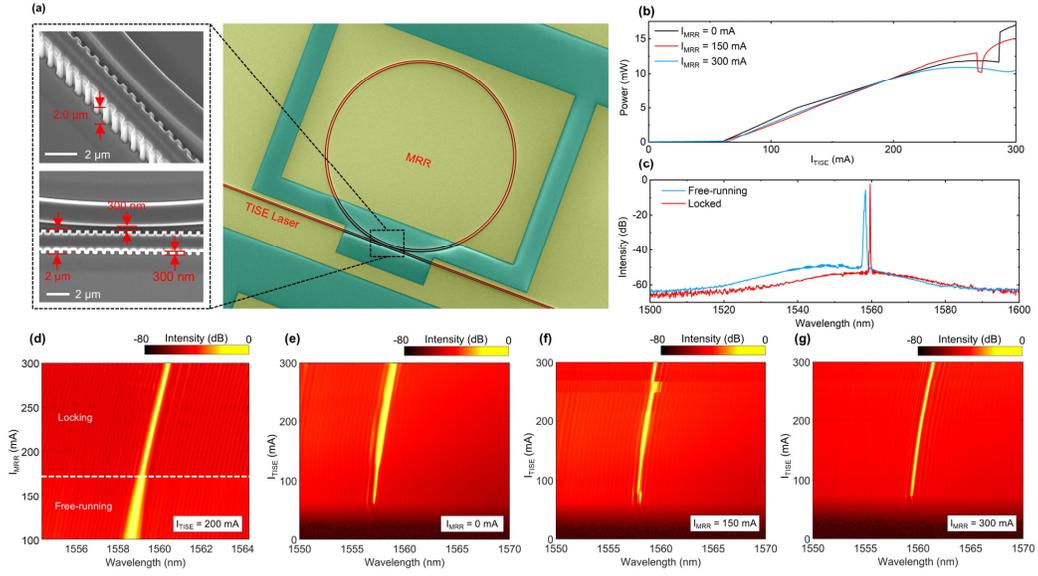

**a** SEM picture of the HF-OIL laser. **b** Typical $I_{TISE}$-$P$ characteristics with different $I_{MRR}$. **c** Optical spectrum with the comparison of free running ($I_{MRR}$ = 0 mA) and locking state ($I_{MRR}$ = 150 mA), $I_{TISE}$ = 300 mA. **d** 2D spectrum as a function of $I_{MRR}$ with $I_{TISE}$ = 200 mA. **e-f** 2D spectrum as a function of $I_{TISE}$ with $I_{MRR}$ = 0 mA, 150mA and 300 mA, respectively.

**Fig. 3: Linewidth and phase noise**

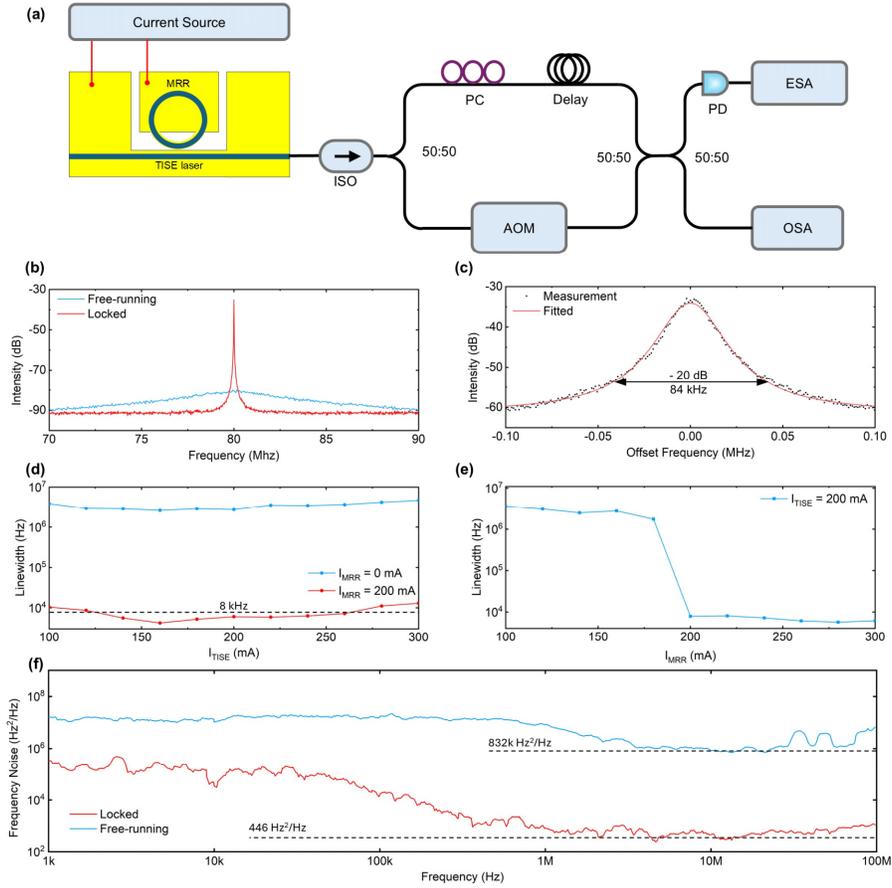

**a** Experimental setup: Light is collected from the TISE facet with a lensed fiber, and non-zero frequency self-heterodyne setup is used to measure the linewidth. ISO, isolator; AOM, acoustic-optic modulator; PC, polarization controller; PD, photodetector; ESA, electrical spectrum analyzer; OSA, optical spectrum analyzer; **b** Measured RF signal measured from PD with injecting locked $I_{MRR}$ = 200 mA and $I_{TISE}$ = 160 mA, yielding a dramatic narrowing of the emission linewidth compared to the free-running regime ($I_{MRR}$ = 0 mA). **c** Voigt fitted linewidth with $I_{MRR}$ = 200 mA and $I_{TISE}$ = 160 mA. **d** linewidth as a function of $I_{TISE}$ with $I_{MRR}$ = 0 mA and 200 mA. **e** linewidth as a function of $I_{MRR}$ with $I_{TISE}$ = 200 mA. **f** Noise spectra of laser measured by correlated delayed self-heterodyne measurement between locked ($I_{MRR}$ = 160 mA and $I_{TISE}$ = 200 mA), and free-running signal ($I_{MRR}$ = 0 mA and $I_{TISE}$ = 200 mA).

**Fig. 4: Injection locking phase encoding**

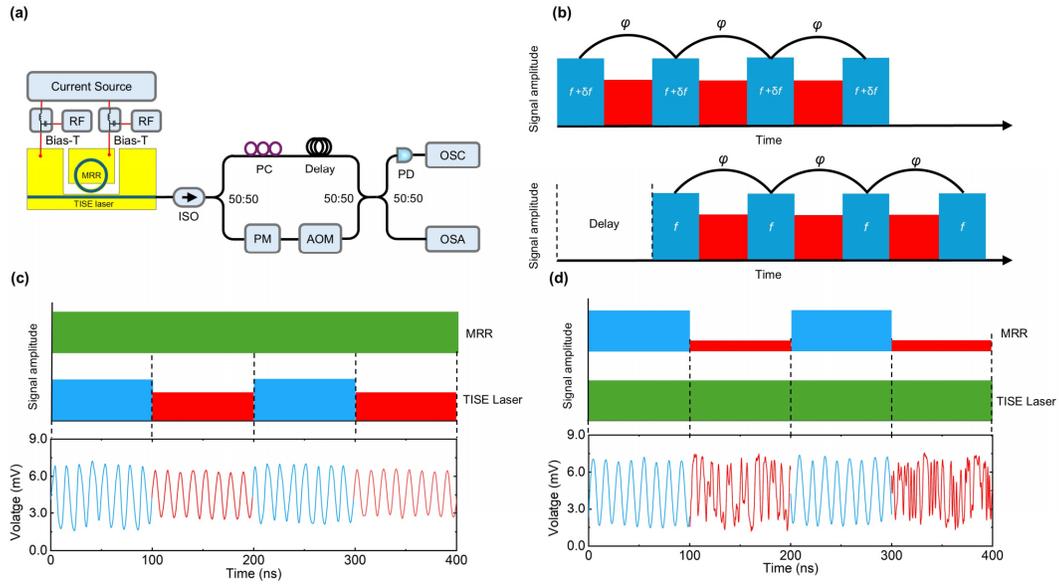

**a** Experimental setup of the AMZI; PM, phase modulator; OSC, oscilloscope. **b** Principle of operation of the phase detection. **c** When the MRR operates in CW and is injected into the TISE laser, the TISE laser pulses all inherit the coherence. **d** When the TISE operates in CW and MRR switches from locking and free-running state, the TISE laser pulses switch between phase-locked and random state.